\documentclass[preprint2]{aastex}
\usepackage{amsmath}
\newcommand{\ud}{\,\mathrm{d}}
\newcommand{\bfOmega}{\,\mathbf{\Omega}}
\newcommand{\bfv}{\,\mathbf{v}}
\newcommand{\bfu}{\,\mathbf{u}}
\newcommand{\bfr}{\,\mathbf{r}}
\newcommand{\bfk}{\,\mathbf{k}}
\newcommand{\hstp}{\,{\frac{\ud s}{2}}}

\begin{document}

\title{Algorithm for Tracing Radio Rays in Solar Corona and Chromosphere}

\author{L. Benkevitch}
\affil{MIT Haystack Observatory, Westford MA}
\and
\author{I. Sokolov}
\affil{AOSS, University of Michigan, Ann Arbor MI}
\and
\author{D. Oberoi}
\affil{MIT Haystack Observatory, Westford MA}
\and
\author{T. Zurbuchen}
\affil{AOSS, University of Michigan, Ann Arbor MI}

\begin{abstract} \small 
In this paper a new efficient algorithm for computation of radio wave ray
trajectories in isotropic media is described. The algorithm is based on an original second-order difference scheme with a specific ``length-conservation'' property, which allows to resolve the ray shape even in the regions where its curvature is high. Besides the scheme, the algorithm includes a number of mechanisms securing its correctness and stability. The algorithm is intended for obtaining multi-pixel solar images in wide range of radio frequencies, and it is designed to be used in studies of the solar environment with modern high-resolution radiointerferometers and radiotelescopes such as the Murchison Widefield Array (MWA). \\
\end{abstract}

\section{Introduction}

The problems of image simulation and reconstruction in solar radioastronomy
require computation of multiple ray trajectories. A typical image of the sun
consists of thousands to millions pixels, and the brightness of each of the
pixels is obtained as an integral over an individual ray trajectory. Due to
non-uniform distribution of the index of refraction in heliospheric plasmas at
lower frequencies (up to a few GHz) the ray trajectories cannot be treated as
straight lines. The ray-tracing problems require significant amount of
computations, and, hence, time. It is possible to considerably reduce the
computation time through the use of efficient ray-tracing algorithms.  

In this work a ray tracing algorithm is presented. It allows one to calculate
trajectories of electromagnetic rays in space plasmas with specified density
distributions. In most applications the radio rays are traced to calculate
certain physical quantities, such as intensity or black body brightness
temperature, through path integration over the ray trajectories. In its software
implementation the algorithm performs the integration and the ray trajectory
calculation simultaneously. 

We start with a brief review of the facts from the physics of electromagnetic
wave propagation in plasmas pertinent to the ray tracing. We show the
relationship between the plasma density, the dielectric permittivity, and the
index of refraction as well as between their gradients. In the next section we
derive the ray equations from the Fermat's principle. We transform the set of ray
equations so that their solution is a naturally parameterized curve. This
representation reveals the mathematical analogy of the ray equations and the
equations for a particle motion in magnetic field. Using this analogy, we derive
a \citet{crn47} second-order difference scheme using the \citet{bor70} approach.
In the Boris' scheme the particle energy, or the squared velocity, is conserved.
For the naturally parameterized curve like that we use here the first derivative
of the vector to be solved from the equation is not velocity (as present in the
equation of the particle motion), but the unity-length direction cosine vector.
Our scheme conserves the squared direction vector at unity. To achieve this, we
employ the ideas which are often used while simulating the particle motion and
which are known to conserve the velocity vector length in the magnetic field.  

The difference scheme coding is quite straightforward, so in the section devoted
to the numerical implementation of the algorithm we focus on discussing the ways
to secure the algorithm robustness for the media with non-linear plasma density
changing in space by presumably decades of orders of magnitude. This is shown to
be the case for the heliosphere. We describe the methods we used to adaptively
decrease and increase the integration step and the techniques used to avoid the
ray penetration into the regions with the density above the critical. These
methods are intended to reverse the ray direction and use parabolic and linear
approximations of the trajectory.

In the testing and validation section we show the algorithm demonstrates
excellent precision in the medium with one-dimensional linear plasma density
gradient. In the example application section we verify the algorithm through
solving a ray-tracing problem in the corona and chromosphere with realistic
model density distributions, those by \citet{sai70} and \citet{cme35}. Our
algorithm implementation shows good results, demonstrated in a few plots of the
ray beams refracting near the sun. 

\section{Fundamentals of ray refraction in isotropic plasmas}

We treat the electromagnetic ray trajectory as a curve with radius vector $\bfr$
in the three-dimensional space. For any starting point, $\bfr_0$, and any given
vector of initial direction, $\bfv_0$, the whole shape of the ray trajectory may
be unamnigously determined, as long as the distribution of the index of
refraction, $n(\bfr,\omega)$, is known for a given angular frequency $\omega$.

For the natural parameterization of the curve, $\bfr(s)$, where $s$ is its arc
length, the direction vector $\bfv(s)$ is defined as the first derivative of
position vector, or $\bfv(s) = \bfr'(s)$. The components of $\bfv(s)$ are
direction cosines, and $|\bfv(s)| = 1$ at any point of the curve. Here we show
that the plasma density distribution, $\rho(\bfr)$, suffices to determine the
index of refraction $n(\bfr)$ at a given wave frequency. In the Gaussian units, $n$ is the square root of plasma dielectric permittivity, $n = \sqrt \varepsilon$.
For isotropic plasmas 
\begin{equation}
  \label{n2_isotropic}
  n^2 = 1 - \frac{\omega^2_p}{\omega^2}, 
\end{equation}
where the plasma frequency is $\omega_p = (4 \pi e^2 n_e / m_e)^{1/2}$, $e$ being the electron charge, $m_e$ and $n_e$ the electron mass and number density in the Gaussian (CGS)  units. Under the quasineutrality condition the hydrogen plasma density is $\rho = m_p n_e$, where $m_p$ is the proton mass, so the dielectric permittivity is expressed as
\begin{equation}
  \label{eps_via_rho}
  \varepsilon =  1 - \frac{\rho}{\rho_{cr}}
\end{equation}

Here $\rho_{cr}$ is the critical plasma density, at which the permittivity and,
hence, the refraction index for the given frequency are zero:
\begin{equation}
  \label{rho_cr}
  \rho_{cr} =  \frac{m_p m_e \omega^2}{4 \pi e^2}
\end{equation}
Electromagnetic rays of the frequency $\omega$ cannot travel in the regions with
$\omega_p > \omega$, i.e. with $\rho > \rho_{cr}$ where the index of refraction
takes imaginary values. The boundary between the region where $\rho < \rho_{cr}$
and the region with $\rho > \rho_{cr}$ is called the critical surface. The rays
near the critical surfaces undergo strong refraction bending them away from the
surface. In limit cases this effect is similar to the reflection. 

The algorithm described further uses the relative gradient of refractive index,
$\nabla n / n$. Since the permittivity gradient can be expressed from Eq.
\eqref{eps_via_rho} as $\nabla \varepsilon = -\nabla \rho / \rho_{cr}$, the
relative gradient of $n$ is
\begin{equation}
  \label{rel_grad_n}
  \frac{\nabla n}{n} = \frac{\nabla \varepsilon}{2 \varepsilon} =
    - \frac{\nabla \rho}{2(\rho_{cr} - \rho)}
\end{equation}
Thus, if we know the critical density, the relative gradient of refractive index
is totally determined by the plasma density and the density gradient.

\section{Ray equation for isotropic media}

To derive the equation for the ray trajectory we use the Fermat's principle. It states that a ray joining two points $P_1$ and $P_2$ will choose the path over which the integral of the refraction index, $n$, reaches its minimum, so that the
variation of the integral about this path is zero:
\begin{equation}
  \label{Fermat}
  \delta \int_{P_1}^{P_2} n(\bfr) \mathrm{d} s
\end{equation}
Notice that the integration in \eqref{Fermat} occurs along a path, while in the
Hamiltonian principle of least action the integration is with respect to the
time. Transform the integrand of \eqref{Fermat} into the form of the Lagrangian 
by changing $s$ for a time-like independent variable $\tau$. If $s = s(\tau)$,
then $\mathrm{d}s^2 = \dot{r}^2 \mathrm{d}\tau^2$, where $\dot{r}^2 = \dot{\bfr}
\cdot \dot{\bfr}$. Substitution of $\mathrm{d}s$ into \eqref{Fermat},
\begin{equation}
  \label{Fermat_tau}
  \delta \int_{P_1}^{P_2} n(\bfr) \sqrt{\dot{r}^2} \ud \tau = 0,
\end{equation}
turns the integrand into the Lagrangian $L(\bfr, \mathbf{\dot{r}}, \tau) =
n(\bfr) \sqrt{\dot{r}^2}$. We apply the Euler method to \eqref{Fermat_tau} to
find the vector differential equation of the curve satisfying the Fermat's
principle. The Euler's equation is
\begin{equation} \label{euler_eq}
  \frac{\ud}{\ud \tau} \left( \frac{\ud L}{\ud \dot{\bfr}} \right) 
      - \frac{\ud L}{\ud \bfr} = 0
\end{equation}
The Lagrangian derivatives with respect to $\dot{\bfr}$ and  $\bfr$ are
\begin{equation} \label{lagr_deriv_rdot}
  \frac{\ud L}{\ud \dot{\bfr}} = 
      n(\bfr) \frac{\dot{\bfr}}{\sqrt{\dot{r}^2}} = 
      n(\bfr) \frac{\dot{\bfr} \ud \tau}{\sqrt{\dot{r}^2} \ud \tau} =
      n(\bfr) \frac{\ud \bfr}{\ud s}
\end{equation}
and
\begin{equation} \label{lagr_deriv_r}
  \frac{\ud L}{\ud \bfr} = \frac{\ud n(\bfr)}{\ud \bfr} \sqrt{\dot{r}^2} = 
      \nabla n(\bfr) \frac{\ud s}{\ud \tau}.
\end{equation}
Therefore, the Euler's equation \eqref{euler_eq} for our problem \eqref{Fermat_tau} takes the form 
\begin{equation} \label{euler_ray_eq}
  \frac{\ud}{\ud \tau} \left(n \frac{\ud \bfr}{\ud s} \right) 
      - \nabla n \frac{\ud s}{\ud \tau} = 0,
\end{equation}
which, after multiplying both parts by ${\ud \tau}/{\ud s}$, yields the ray equation:
\begin{equation} \label{ray_eq_0}
  \frac{\ud}{\ud s} \left(n \frac{\ud \bfr}{\ud s} \right) - \nabla n = 0.
\end{equation}
With the use of the vector identities, Eq. \eqref{ray_eq_0} can be represented
in the format with explicit first and second derivatives of $\bfr$ as
\begin{equation}
  \label{ray_eq}
  \frac{\ud^2 \bfr}{\ud s^2} = \frac{\ud \bfr}{\ud s} \times 
      \left(\frac {\nabla{n}} {n} \times \frac{\ud \bfr}{\ud s} \right)
\end{equation}
This is a system of three second-order equations. By introducing the direction
vector $\bfv = \ud \bfr / \ud s$ the system \eqref{ray_eq} is transformed into
the set of six first-order equations
\begin{equation}
  \label{ray_eq_6sys}
  \left\{
    \begin{aligned}
      \frac{\ud \bfr}{\ud s} &= \bfv \\
      \frac{\ud \bfv}{\ud s} &= \bfv \times \left(\frac {\nabla{n}} {n} \times
\bfv \right)
    \end{aligned}
  \right.
\end{equation}
The set of equations \eqref{ray_eq_6sys} is numerically solved by the proposed
ray-tracing algorithm. Note that the independent variable $s$ is the ray arc
length so the solution to \eqref{ray_eq_6sys} is a naturally parameterized curve
$\bfr = \bfr(s)$. This infers the ``conservation law'' for $\bfv(s) = \bfr'(s)$,
\begin{equation} \label{abs_v_equ_1}
  |\bfv| \equiv 1,
\end{equation}
at any point of the curve. The algorithm of ray tracing described in
Section~\ref{sec_alg_descr} is based on a scheme that conserves this property.

\section{Ray equation as particular case of Haselgrove equations}

It is intrersting to observe that equation \eqref{ray_eq_0} is a particular case of a more general set of ray equations derived by \citet{wal08}:
\begin{equation}
  \label{walker_ray_eq}
  \left\{
    \begin{aligned}
      \frac{\ud \bfr}{\ud t} &= \nabla_k \omega \left(\bfr,\bfk \right) \\
      \frac{\ud \bfk}{\ud t} &= - \nabla \omega \left(\bfr,\bfk \right) \\
      &\!\!\!\!\!\!\!\!  G \left( \omega, \bfk \right) = 0
    \end{aligned}
  \right.
\end{equation}
In Eqs. \eqref{walker_ray_eq} $G \left( \omega, \bfk \right) = 0$ is the dispersion relation of the medium, $t$ is time, ${\ud \bfr}/{\ud t} = \dot{r} = \mathbf{V}_G$ is the group velocity of the wave packet, $\omega$ is the angular frequency, $\bfk$ is the wave vector, and $\nabla_k$ is the differentiation operator with respect to the $\bfk$ components. System \eqref{walker_ray_eq} is a modification of the \citet{has55} equations. Its solutions are trajectories of rays of virtually any nature, not just electromagnetic, as long as the waves are plane and linear, varying in space and time as $\mathrm{exp}[i(\bfk \cdot \bfr - \omega t )]$. 

Let us show that equation \eqref{ray_eq_0} is equivalent to system \eqref{walker_ray_eq} for the waves traveling through a collisionless isotropic plasma with the dispersion relation
\begin{equation}
  \label{disp_rel}
  {\omega}^2 = c^2 k^2 - {\omega_p }^2,
\end{equation}
which is easily found from equation \eqref{n2_isotropic} and the relation between the index of refraction and the magnitude $k$ of the wave vector, 
\begin{equation}
  \label{n_is_ck_over_omega}
  n = c k / \omega.
\end{equation}
Hereafter $c$ denotes the speed of light in the vacuum. In equations \eqref{walker_ray_eq} we change variables, time $t$ for arc length $s$. For a curve $\bfr = \bfr(t)$, where $t$ is a parameter, the element of arc length 
\begin{equation}
  \label{ds_is_vgdt}
  \ud s = |\dot{r}| \ud t = |\mathbf{V}_G| \ud t.
\end{equation}
Differentiation of \eqref{disp_rel} with respect to $\bfk$ gives the first equation of system \eqref{walker_ray_eq}:
\begin{equation}
  \label{walker_1st}
  \frac{\ud \bfr}{\ud t} = \nabla_k \omega = \frac{c^2}{\omega} \bfk, 
\end{equation}
and therefore the group velocity and its absolute value are
\begin{equation}
  \label{nabla_k_omega}
  \mathbf{V}_G = \nabla_k \omega = \frac{c^2}{\omega} \bfk, 
    \qquad V_G = \frac{c^2}{\omega} k.
\end{equation}
Then, according to \eqref{ds_is_vgdt}, $\ud t$ is rendered as follows:
\begin{equation}
  \label{arclen_via_dt}
  \ud t = \frac{\omega}{c^2 k} \ud s. 
\end{equation}
Substitution of $\ud t$ in equation \eqref{walker_1st} yields
\begin{equation}
  \label{k_over_k}
  \frac{\ud \bfr}{\ud s} = \frac{\bfk}{k}.
\end{equation}
The second equation of \eqref{walker_ray_eq} is obtained in a similar way by applying $\nabla$ to the dispersion relation \eqref{disp_rel},
\begin{equation}
  \label{walker_2nd}
  \frac{\ud \bfk}{\ud t} = - \nabla \omega = - \frac{\omega_p}{c^2 k} 
      \nabla \omega_p, 
\end{equation}
and using \eqref{arclen_via_dt} to change $\ud t$ for $\ud s$:
\begin{equation}
  \label{dk_over_ds}
  \frac{\ud \bfk}{\ud s} = - \frac{\omega_p}{c^2 k} \nabla \omega_p.
\end{equation}
We reduce equation \eqref{k_over_k} to a form similar to that of \eqref{ray_eq_0}. Multiply both sides by $c k/\omega$ and use \eqref{n_is_ck_over_omega}:
\begin{equation}
  \label{n_by_drds}
  n \frac{\ud \bfr}{\ud s} = \frac{c}{\omega} \bfk.
\end{equation}
Differentiate both sides with respect to $\ud s$:
\begin{equation}
  \label{diff_n_by_drds}
  \frac{\ud}{\ud s} \left( n \frac{\ud \bfr}{\ud s} \right) = \frac{c}{\omega} 
      \frac{\ud \bfk}{\ud s}.
\end{equation}
Substitution of \eqref{dk_over_ds} in the right-hand-side yields
\begin{equation}
  \label{diff_n_by_drds2}
  \frac{\ud}{\ud s} \left( n \frac{\ud \bfr}{\ud s} \right) = - \frac{\omega_p}{\omega c k} \nabla \omega_p.
\end{equation}
The left-hand-side of \eqref{diff_n_by_drds2} is the same as the first term in \eqref{ray_eq_0}, while its right-hand-side is $\nabla n$. This can be verified by applying $\nabla$ operator to equation \eqref{n2_isotropic} and using the dispersion relation \eqref{disp_rel}. Thus we have shown that equation \eqref{ray_eq_0} derived directly from the Fermat's principle in the previous Section is equivalent to the general ray equations \eqref{walker_ray_eq} constrained by the conditions of plasma isotropy and absence of losses.

\section{Algorithm description} \label{sec_alg_descr}

The integration of system \eqref{ray_eq_6sys} is based on an original algorithm
developed as a modification of the Boris' implementation of the explicit
\citet{crn47} scheme for the calculation of a charged particle motion in
magnetic field \citep{hoc88}. The \citet{bor70} CYLRAD algorithm secures the
energy conservation, which is equivalent to conservation of the particle's
squared velocity $\bfv^2$. The ray trajectory calculations require a different
(though similar) conservation law to be observed. 

Below is shown a brief derivation of the second-order difference scheme for the
system \eqref{ray_eq_6sys}. We introduce a vector quantity $\bfOmega$ as
\begin{equation} \label{omega}
  \mathbf \Omega = \frac {\nabla n} {n} \times \mathbf v
\end{equation}
 Substitute \eqref{omega} into \eqref{ray_eq_6sys} to render the ray equations
in the form
\begin{equation} \label{ray_eq_omega}
  \left\{
    \begin{aligned}
      \frac{\ud \bfr}{\ud s} &= \bfv \\
      \frac{\ud \bfv}{\ud s} &= \bfv \times \bfOmega.
    \end{aligned}
  \right.
\end{equation}
Formally, this set of equations is the same as (4-90) in the book by
\citet{hoc88}. However, in the ray trajectory equations \eqref{ray_eq_omega} the
independent variable $s$, the ray arc length, replaces the time $t$, introducing
the natural ray parameterization. The ray direction $\bfv(s) = \bfr'(s)$ is used
instead of the particle velocity $\bfv(t) = \bfr'(t)$, and the vector
$\bfOmega(s)$ is used instead of the cyclotron frequency $\mathbf \Omega = q
\mathbf{B} / m$ . Analogously, the square of direction vector $\bfv^2(s) = 1$ 
must also be conserved in the course of numerical integration, because
$\bfv^2(s) = \bfr'^2(s) \equiv 1$ for any regular point of a naturally
parameterized curve. 

It is easy to show that $\bfOmega$ is a vector whose length equals the ray
curvature, defined as $\kappa = |\bfr''(s)|$ for naturally parameterized curves.
Render the second equation in \eqref{ray_eq_omega} as $\bfr'' = \bfr' \times
\bfOmega$. Multiply both sides of it by $\bfr' \times$ on the left:
\begin{equation}
  \bfr' \times \bfr'' = \bfr' \times (\bfr' \times \bfOmega)
\end{equation}
Expanding the right-hand-side with the use of the vector identity and keeping in
mind that $\bfr' \cdot \bfr' = 1$ and $\bfr' \perp \bfOmega$, we get
\begin{equation}
  \bfOmega = - \bfr' \times \bfr''
\end{equation}
But for the natural parameterization $\bfr' \perp \bfr''$, hence, $|\bfOmega| =
|\bfr''|$, and 
\begin{equation} \label{curv}
  |\bfOmega| = \kappa
\end{equation}

 The numerical solution is sought at the nodes of one-dimensional grid marked on
the ray trajectory. Consider a grid whose nodes with respect to some $i$-th
position are numbered as ... $(i-1)$, $(i-1/2)$, $(i)$, $(i+1/2)$, $(i+1)$,...
and so forth. We subscript a variable with 0 if its value is taken at $i$-th
point, with $(1/2)$ for $(i+1/2)$-th point, and with 1 for $(i+1)$-th point. A
“leapfrog” difference scheme is used, where the half-step points are used in
calculations. Suppose a ray at a particular moment has its end position $\bfr_0$
and direction $\bfv_0$. The ray tracing scheme provides formulae to calculate
its successive position $\bfr_1$ and direction $\bfv_1$ after the step $\ud s$
along the ray arc length. The variables subscripted with $1/2$ are intermediate
and they are related to the middle of the step. Note that the density and its
gradient are provided only once for each step at the step middle point so the
$(\nabla n / n)$ ratio is ascribed to the grid point $(i+1/2)$. The second
differential equation in \eqref{ray_eq_omega} can be approximated by the
differences as
\begin{equation}
  \frac{\bfv_1 - \bfv_0}{\ud s} = 
    \bfv_{1/2} \times \bfOmega_{1/2}
\end{equation}
Approximation of $\bfv_{1/2}$ as $\frac{1}{2}(\bfv_1 + \bfv_0)$ yields the
implicit \citet{crn47} scheme
\begin{equation} \label{impl_crn}
  \bfv_1 - \bfv_0 = (\bfv_1 + \bfv_0) \times
    \bfOmega_{1/2} \hstp
\end{equation}
This scheme has the benefit of unconditional stability. Multiply both sides of
\eqref{impl_crn} scalarly by $\bfOmega_{1/2}$ to get the property:
\begin{equation} \label{vomega}
  \bfv_1 \cdot \bfOmega_{1/2} = \bfv_0 \cdot \bfOmega_{1/2}, 
\end{equation}
which will be used later. To convert scheme \eqref{impl_crn} into the explicit
one, express $\bfv_1$ from \eqref{impl_crn} as
\begin{equation}
  \bfv_1 = \bfv_0 + (\bfv_1 + \bfv_0) \times
    \bfOmega_{1/2} \hstp
\end{equation}
and substitute it in the right-hand-side of \eqref{impl_crn} for itself. It yields 
\begin{align}
  \bfv_1 &= \bfv_0 
    + 2 \bfv_0 \cdot \bfOmega_{1/2} \hstp \nonumber \\ &
    + \left( (\bfv_1 + \bfv_0)
    \times \bfOmega_{1/2} \hstp \right) 
    \times \bfOmega_{1/2} \hstp
\end{align}
Using the identity $\bfv_0 = \bfv_0 (1 + (\bfOmega_{1/2} \hstp)^2 -
(\bfOmega_{1/2} \hstp)^2)$, vector identities, property \eqref{vomega}, and
rearranging the terms, we arrive at the explicit difference scheme known as the
Boris' CYLRAD algorithm \citep{bor70}:
\begin{align}  \label{cylrad}
  \bfv_1 &= \bfv_0 + 
    \frac{2}{1 + \left(\bfOmega_{1/2} \hstp\right)^2} \nonumber \\ &
    \left(\bfv_0 + 
    \bfv_0 \times \bfOmega_{1/2} \hstp \right)
    \times \bfOmega_{1/2} \hstp
\end{align}
The vector $\bfOmega_{1/2}$ requires extrapolation: as seen from \eqref{omega},
the first multiplier, ${\nabla n}/n$, is obtained at the intermediate point and
therefore has the $(1/2)$ subscript, while the second, $\bfv$, is only available
for the starting point subscripted with 0. We denote this cross-product as
$\bfOmega_0$,
\begin{equation} \label{omega_0}
  \bfOmega_0 = \left( \frac{\nabla n}{n} \right)_{\frac{1}{2}} 
    \times \bfv_0,
\end{equation}
and use it as an initial value for $\bfOmega$ extrapolation up to
$\bfOmega_{1/2}$. The half-step difference scheme for the second equation in
\eqref{ray_eq_omega} is
\begin{equation}
  \bfv_{1/2} = \bfv_0 + \bfv_0 \times \bfOmega_0 {\hstp}
\end{equation}
Substitution of $\bfv_{1/2}$ into \eqref{omega} yields the extrapolated
$\bfOmega$:
\begin{equation}  \label{omega_12}
  \bfOmega_{1/2} = \left( \frac{\nabla n}{n} \right)_{\frac{1}{2}} 
    \times \left( \bfv_0 + \bfv_0 \times \bfOmega_0 {\hstp} \right)
\end{equation}
In order to make the algorithm faster we reduce the number of multiplications.
Note that all the occurrences of $\bfOmega$ in \eqref{cylrad} and
\eqref{omega_12} have the multiplier ${\ud s}/2$, so it is expedient to include
it in the expressions for $\bfOmega$ in  \eqref{omega_0} and \eqref{omega_12}.
Finally, we put together the formulae \eqref{omega_0}, \eqref{omega_12}, and
\eqref{cylrad} and add the position vector $\bfr$ adjustments at the start and
the end to form the ray tracing algorithm \eqref{alg}. 

The algorithm presented below converts the initial ray position and direction
$(\bfr_0,\bfv_0)$ into the new ones $(\bfr_1,\bfv_1)$ along the ray path in five
steps:
\begin{subequations} \label{alg}
\begin{align} 
  1.\; &\bfr_{1/2} = \bfr_0 + \bfv_0 {\hstp} \label{alg_1}\\
  2.\; &\bfOmega_0 = \left( \frac{\nabla n}{n} \right)_{\frac{1}{2}} 
    \times \bfv_0 \hstp \label{alg_2} \\
  3.\; &\bfOmega_{1/2} = \left( \frac{\nabla n}{n} \right)_{\frac{1}{2}} 
    \times \left( \bfv_0 + \bfv_0 \times \bfOmega_0 \right) \hstp \label{alg_3}\\
  4.\; &\bfv_1 = \bfv_0 + 
    \frac{2}{1 + \left(\bfOmega_{1/2} \hstp\right)^2} \nonumber \\ &
    \hspace{8mm} \left(\bfv_0 + 
    \bfv_0 \times \bfOmega_{1/2} \right)
    \times \bfOmega_{1/2} \label{alg_4} \\
  5.\; &\bfr_{1} = \bfr_{1/2} + \bfv_1 \hstp \label{alg_5} 
\end{align}
\end{subequations}

The first step \eqref{alg_1} provides the $\bfr$ half-step-ahead approximation
$\bfr_{1/2}$. Next two steps \eqref{alg_2} and \eqref{alg_3} calculate the
half-step-ahead approximation for $\bfOmega$. The fourth step \eqref{alg_4}
estimates the new ray direction, $\bfv_1$, based on the previous direction,
$\bfv_0$, and $\bfOmega_{1/2}$. Step five, \eqref{alg_5}, produces the new ray
position, $\bfr_{1}$, through the half-step from the intermediate point
$\bfr_{1/2}$ along the new direction $\bfv_1$.

\section{Numerical implementation}

The optical properties of the media where tracing the electromagnetic rays is of
interest are often extremely nonuniform. For example, the electron number
density $N_e$ within the chromosphere alone falls off with the height by more
than 10 orders of magnitude. Inside the solar corona $N_e$ decays with the solar
distance also by decades of magnitude orders. The index of refraction, as the
square root of proportional to $N_e$ dielectric permittivity
\eqref{eps_via_rho}, can change by many orders of magnitude within the region of
study. The non-linear dependence of the refractive index on the wavelength poses
other issues. For different wavelengths the ray geometry can be very different
because the regions with the density greater than critical \eqref{rho_cr} form
different configurations in space. The numerical implementation of algorithm
therefore must include the mechanisms for adaptive step size $\ud s$ adjustment
not only to ensure the specified precision but also to prevent the ray from
penetration into the region with $\rho > \rho_{cr}$. In some cases $\bfr_0$
appears so close to the region with the critical density that the step $\ud s$
must be shorter than the absolute minimum step specified. For this condition a
special method of replacing the ray trajectory by a segment of parabola or even
straight line is used. All these techniques are described in this Section.

In our implementation of the ray tracing algorithm, aimed at obtaining multi-pixel images, many rays are processed simultaneously. Over a single integration step each of the rays is tested for precision and correctness, and only those that passed the tests are submitted to the Boris's algorithm to calculate new positions and directions. The rays for which the absolute minimum step leads to penetration to the region with $\rho > \rho_{cr}$ are switched to the opposite parabola branches or just reflected from the critical surface. The rest are the rays whose step sizes are too large to secure the specified precision. The step size for them is shortened by applying formula \eqref{ds_corr}. The rays with reduced step sizes are  processed at the next integration step. If a ray is leaving the dense region, its step size $\ud s$ is increased over the series of subsequent integration steps.

The plasma density, $\rho$, and its gradient, $\nabla \rho$, are calculated
through calls to the specialized subroutine once per each algorithm step. It is assumed that these plasma parameters are obtained through an MHD simulation and they are available at the nodes of a three-dimensional grid with variable grid step. The same subroutine provides the estimate for the maximum ray step size consistent with the MHD grid size for each ray. 

\subsection{Precision control through adaptive step size}

The desired precision of the ray trajectory is specified with the variable
$Tol$. It has the meaning of inverse number of the steps $n_s$ per a radian of
the ray arc length. For example, $Tol < 0.1$ means that at least 10 steps per
one radian of ray arc length is required. At each algorithmic step the real
precision is calculated and compared with $Tol$, and if it exceeds $Tol$, the
step $\ud s$ is decreased. 

The curvature $\kappa$ is defined as the inverse of instant curvature radius as
$\kappa = 1/R_c$. The actual ray curvature is available at each algorithm step
as $|\bfOmega|$ due to \eqref{curv}. Instead of the real precision it is
convenient to use the quantity $\tau^2$, which is actually a halved real
precision:
\begin{equation} \label{tausqr}
  \tau^2 = \left( \bfOmega \hstp \right)^2 = \left( \kappa \hstp \right)^2,
\end{equation}
The algorithm step \eqref{alg_4} requires calculation of $\tau^2$ anyway, so
using it saves time. It is $\tau^2$ that is compared with $Tol^2$ to make
decision on the $\ud s$ reduction.  Comparing the halved real precision with the
desired precision $Tol$ makes the step ajustment less frequent as the specified
precision is secured over most of the ray trajectory portions. To reduce the
number of floating-point calculations, the $\tau^2$ approximation uses the
condition $\bfv^2 = 1$ in the identity $(\bfu \times \bfv)^2 = \bfu^2 \bfv^2 -
(\bfu \cdot \bfv)^2 = \bfu^2 - (\bfu \cdot \bfv)^2$. The index of refraction $n$
in \eqref{omega} is replaced by the dielectric permittivity as defined in
\eqref{rel_grad_n}, so
\begin{equation}
  \tau^2 = \left(\hstp \left(
    \frac{\nabla \varepsilon_{1/2}}{2 \varepsilon_{1/2}} \right) \times
    \bfv_0 \right)^2,
\end{equation}
which with the use of the identity yields
\begin{equation}
  \tau^2 = \left(\hstp\right)^2 \left[ \left(
    \frac{\nabla \varepsilon_{1/2}}{2 \varepsilon_{1/2}} \right)^2 -
    \left(\frac{\nabla \varepsilon_{1/2}}{2 \varepsilon_{1/2}} \cdot
    \bfv_0 \right)^2 \right]
\end{equation}
The step must be $\ud s = R_c / n_s = Tol / \kappa$ to have exactly $1/Tol$
points per radian of the ray, so due to \eqref{tausqr} it is corrected as:
\begin{equation} \label{ds_corr}
  \ud s' = \frac{Tol}{\sqrt{\tau^2}} \hstp,
\end{equation}

The subroutine that provides the plasma density and density gradient at each
call also provides the desired new step size $\ud s_{new}$ for each ray. After
the positions and directions for the rays have been advanced with the current
step sizes, the algorithm attempts to increase the steps towards $\ud s_{new}$.
However, $\ud s_{new}$ can be greater than the current $\ud s$ by orders of
magnitude, and near the critical surface a simple assignment $\ud s = \ud
s_{new}$ would almost inevitably lead to the ray penetration past the critical
surface. Therefore the algorithm uses a special method for gradual increase of
the step size. It is based on running a non-linear difference scheme which
outputs the increased $\ud s'$ from the current $\ud s$ and the desired $\ud
s_{new}$.
\begin{figure}[h]
\plotone{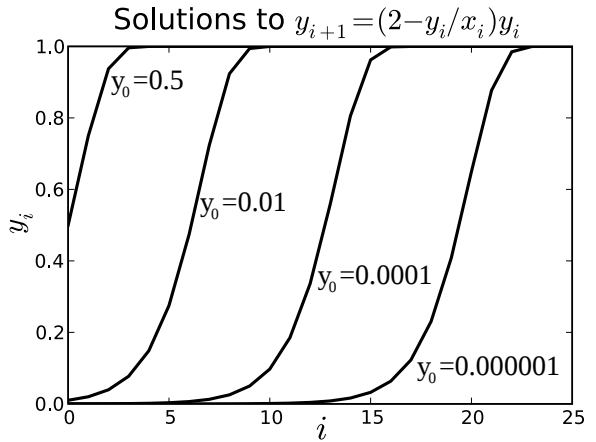}
\caption{\small Solutions to the non-linear difference equation 
  \eqref{nlineq} for four different initial conditions $y_0$. The input variable
is constant: $x_i = 1$. The solutions, $y_i$, are increasing functions, always
converging to $x_i = 1$. This scheme is used to smoothly increase the step $\ud
s$ to the set value of $\ud s_{new}$. \label{nlinstepincr_f1}}
\end{figure}

Consider a difference equation associating two variables, the input x and the
output y, specified at the points ... (i-1), (i), (i+1), (i+2) ... and so on:
\begin{equation} \label{nlineq}
  y_{i+1} = (2 - \frac{y_i}{x_i})y_i
\end{equation}
For constant input $x_i = X$ the solution to Eq. \eqref{nlineq} is stable for
any initial condition on the output variable $y_0$ between 0 and $X$. The
solution to Eq. \eqref{nlineq} always converges to the value $X$. The useful
property of Eq. \eqref{nlineq} is that for a small (compared to $X$) initial
condition $y_0$ the solution behaves similar to growing exponential function
$e^t$, while for large values of $y_0$, i.e. those closer to the X, it grows
slower and slower similar to the decaying exponent $Xe^{-t}$, eventually
converging to $X$. This behavior is illustrated in Fig. \ref{nlinstepincr_f1}.
Notice that reasonably large initial values for $y_0$ ensure fast convergence
within 1\% of the range $[0..1]$: 10 steps for $y_0 = 0.01X$, and even 3 steps
for $y_0 = 0.5X$. 

In our algorithm implementation the step $\ud s$ is incremented using the Eq.
\eqref{nlineq} as the algorithm 
\begin{equation} \label{correq}
  \ud s' = \left(2 - \frac{\ud s}{\ud s_{new}} \right)\ud s
\end{equation}
When $\ud s$ is small compared to $\ud s_{new}$, each next value of $\ud s'$ is
almost $2\ud s$, so $\ud s$ grows in a geometrical progression. However, as ds
approaches$\ud s_{new}$, its growth rate becomes slower. These features ensure
fast and yet smooth adjustment of $\ud s$ in several steps up to the specified
$\ud s_{new}$ value. 

\subsection{Correctness control}

Correctness of the ray path computation is violated if at some step the ray
penetrates the critical surface. This condition can arise for very steep rays,
almost normally incident on the critical surface. To prevent a ray from
trespassing the critical surface the algorithm makes a linear prediction on the
next value of dielectric permittivity $\varepsilon$.  If the predicted
$\varepsilon$ is negative, the ray is approximated by a parabola and it is then
switched to the symmetrical point on the opposite branch of the parabola. This
is similar to the reflection at the critical surface, after which the ray starts
traveling away from it. However, the linear $\varepsilon$ prediction may fail
due to highly non-linear plasma density distribution and then the ray trespasses
the critical surface. This unphysical condition is corrected by returning the
ray its previous position and direction, numerical calculation of the distance
to critical surface along this direction, and linear reflection of the ray at
the critical surface.
  
When the current ray state is $(\bfr_0,\bfv_0)$ and the algorithm is calculating
its next position and direction $(\bfr_1,\bfv_1)$, the mid-step values of the
plasma density $\rho_{1/2}$ and its gradient $\nabla \rho_{1/2}$ are obtained
from the subroutine, so the dielectric permittivity $\varepsilon_{1/2} > 0$ and
its gradient $\nabla \varepsilon_{1/2}$ are available from \eqref{eps_via_rho}
and \eqref{rel_grad_n}, respectively. If the ray at next step  occasionally
trespasses the critical surface, the next value of permittivity,
$\varepsilon_{3/2}$, will be negative. Therefore, predicting the sign of
$\varepsilon_{3/2}$ helps the algorithm to make decision whether to use the
Boris' scheme \eqref{alg} or take measures to avoid the penetration past the
critical surface. The prediction is based on testing the condition
\begin{equation} \label{condeps}
  \varepsilon_{3/2} / \varepsilon_{1/2} > 0
\end{equation}

Replace both numerator and denominator in \eqref{condeps} by the linear portions
of their Taylor series:
\begin{align} \label{epstaylor}
  \frac{\varepsilon_{3/2}}{\varepsilon_{1/2}} &=  
    \frac
      {\varepsilon_0 + \bfv_0 \cdot \nabla \varepsilon_{1/2} 3\hstp}
      {\varepsilon_0 + \bfv_0 \cdot \nabla \varepsilon_{1/2} \hstp} 
    \nonumber \\ &= 1 + \frac
      {\bfv_0 \cdot \nabla \varepsilon_{1/2}}
      {\varepsilon_0 + \bfv_0 \cdot \nabla \varepsilon_{1/2} \hstp} 
	\ud s > 0,
\end{align}
where $\varepsilon_0$ is the permittivity value at the point $\bfr_0$. After
rearranging the terms and dividing both parts by $\varepsilon_0$ we arrive at
the correctness criterion:
\begin{equation}  \label{correctness}
  \hstp \left(\bfv_0 \cdot 
    \frac{\nabla \varepsilon_{1/2}}{\varepsilon_0} \right) > 
      -\frac{1}{3}
\end{equation}
Since the plasma density subroutine provides only the half-step value,
$\rho_{1/2}$ and, hence, only the half-step permittivity value,
$\varepsilon_{1/2}$, can be directly calculated, $\varepsilon_0$ is obtained as
its back approximation using the linear terms of its Taylor series:
\begin{equation} \label{eps0}
  \varepsilon_0 = \varepsilon_{1/2} - 
    \bfv_0 \cdot \nabla \varepsilon_{1/2} \hstp
\end{equation}
If inequality \eqref{correctness} is true, the next step is supposed to be safe.
If not, the ray is so close to the critical surface and so steep that it will
penetrate the critical surface at the current step $\ud s$. In this case the
next position and direction are calculated through parabolic approximation.

\subsubsection{Parabola switching}

\begin{figure}[h]
\plotone{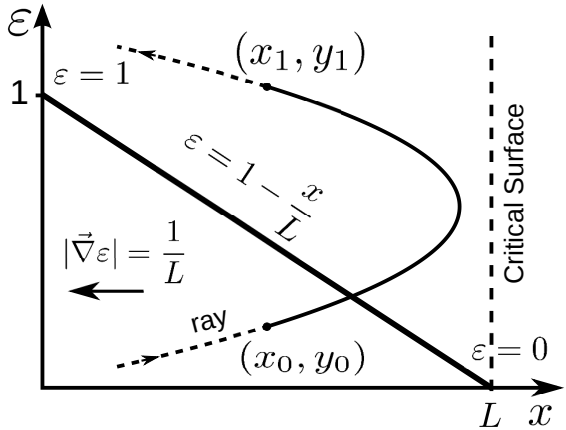}
\caption{\small In a medium with linearly changing dielectric permittivity
  $\varepsilon$ electromagnetic ray trajectories are second-order parabolae. The
$\varepsilon$ linearly changes from 1 at $x = 0$ to 0 at $x = L$. The critical
surface is the locus where $\varepsilon = 0$. \label{rayparabl_f2}}
\end{figure}

This method is used when the correctness criterion \eqref{correctness} fails and
the ray is about to trespass the critical surface. If the step $\ud s$ is small
enough, the dielectric permittivity gradient $\nabla \varepsilon$ can be
considered constant over the ray path. In the medium with $\nabla \varepsilon =
const$ the electromagnetic rays have parabolic trajectories, which will be shown
further. Knowing the parameters of the parabola (see Fig. \ref{rayparabl_f2}),
the algorithm calculates the position and direction vectors $(\bfr_1,\bfv_1)$ of
the ray at the symmetric point of the opposite branch of the parabola. The ray
is switched to the opposite parabolic trajectory branch so that instead of
approaching the critical surface it starts departing it. In order to find the
new position and direction after the parabola switching we solve the ray Eq.
\eqref{ray_eq_0} in two dimensions in the $XY$-plane containing the vectors
$-\nabla \varepsilon_0$ and $\bfv_0$, the $X$ axis collinear with $-\nabla
\varepsilon_0$.

\begin{figure}[h]
\plotone{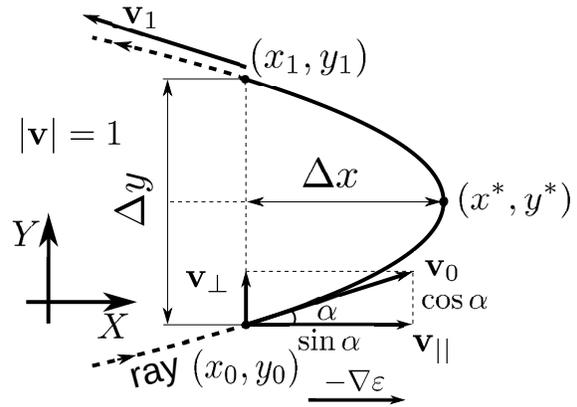}
\caption{\small Parameters of a ray trajectory approximation by the quadratic
  parabola in the $XY$ plane. The $X$ axis is aligned with the negated
dielectric permittivity gradient, $-\nabla \varepsilon$. The direction cosines'
vector $\bfv_0$ at the point $(x_0,y_0)$ is decomposed into two components,
$\bfv_{||}$ along the $X$ and $\bfv_{\perp}$ along the $Y$, with the lengths
equal to $\cos \alpha$ and $\sin \alpha$, respectively. When the ray leaves the
parabola at the point $(x_1,y_1)$, its $x$ coordinate remains the same, but $y$
changes by $\Delta y$, equal to the doubled distance from $y_0$ to the parabola
vertex $y^*$ along the $Y$ axis.  \label{parrayprm_f3}}
\end{figure}

Consider a simple 2-dimensional case of a ray traveling within the $XY$-plane,
where the dielectric permittivity changes linearly from $\varepsilon = 1$ at the
origin of coordinates down to $\varepsilon = 0$ at some distance $L$ along the
$X$ axis, so that
\begin{equation} \label{lineps}
  \varepsilon = 1 - \frac{x}{L}
\end{equation}
The case schematic is shown in Fig. \ref{rayparabl_f2}. We shall derive the
equation for the ray trajectory passing through an arbitrary point $(x_0,y_0)$.
The equation set for the ray trajectory \eqref{ray_eq_0} in the two-dimensional
geometry consists of two equations for $x$ and $y$. The dielectric permittivity
$\varepsilon$ changes in the $X$ direction only, so does the index of refraction
$n$, therefore its gradient component along the $Y$ axis, $\partial n / \partial
y$, is zero, and the equation for $y$ takes the form
\begin{equation}
  \frac{\ud}{\ud s} \left(n \frac{\ud y}{\ud s}  \right) = 0,
\end{equation}
which immediately implies
\begin{equation} \label{n_dyds_is_const}
  n \frac{\ud y}{\ud s} = const
\end{equation}
Refer to the schematic in Fig. \ref{parrayprm_f3}. Denote the values of $n$ and
$\varepsilon$ at the point $(x_0,y_0)$ as $n_0 = \sqrt{\varepsilon_0}$, where
$\varepsilon_0$ is calculated as in \eqref{eps0}. The direction vector component
$\ud y / \ud s = \sin \alpha$, where $\alpha$ is the angle between the ray
direction $\bfv_0$ and $-\nabla \varepsilon_0$, which is the $X$ axis here.
Therefore, the constant in \eqref{n_dyds_is_const} can be presumed in the form
$n_0 \sin \alpha$, so
\begin{equation}
  \frac{\ud y}{\ud s} = \frac{n_0}{n} \sin \alpha
\end{equation}
Using the relationship $\ud x^2 / \ud s^2 + \ud y^2 / \ud s^2 = 1$ and excluding
$\ud s$, we arrive at the differential equation
\begin{equation} \label{diffeqeps}
  \frac{\ud x}{\sqrt{\varepsilon - \varepsilon_0 \sin^2 \alpha}} = 
    \frac{\ud y}{\sqrt{\varepsilon_0} \sin \alpha} 
\end{equation}

Since $\varepsilon$ is a linear function \eqref{lineps}, $\varepsilon(x) =
\varepsilon_0 + (x - x_0)\varepsilon'_0$, where $\varepsilon'_0 = \ud
\varepsilon / \ud x |_{x=x_0}$, so Eq. \eqref{diffeqeps} can be integrated as
\begin{equation} 
  \int_{x_0}^{x} \frac{\ud x}{\sqrt{\varepsilon'_0 x + 
    \varepsilon'_0 x_0 - \varepsilon_0 \cos^2 \alpha}} = 
    \int_{y_0}^{y}\frac{\ud y}{\sqrt{\varepsilon_0} \sin \alpha} 
\end{equation}
to yield the solution in the form $x = x(y)$:
\begin{align} \label{pareq}
  x &= x_0 - \frac{\varepsilon_0}{\varepsilon'_0} \cos^2 \alpha + \nonumber \\ &
    \frac{1}{\varepsilon'_0} \left( \sqrt{\varepsilon_0} \cos \alpha +
    \frac{\varepsilon'_0}{2 \sqrt{\varepsilon_0} \sin \alpha} 
      (y - y_0) \right)^2
\end{align}
This second order parabola equation describes the electromagnetic ray trajectory
for linearly distributed $\varepsilon$. Its vertex point $(x^*,y^*)$ coordinates
are
\begin{subequations} \label{parvert}
\begin{align} 
  y^* & = y_0 - \frac{2 \varepsilon_0 }{\varepsilon'_0 } \cos \alpha \sin
    \alpha \label{y_extr}\\
  x^* & = x_0 - \frac{\varepsilon_0 }{\varepsilon'_0 } \cos^2 \alpha 
    \label{x_extr}
\end{align} 
\end{subequations}

\begin{figure}[h]
\plotone{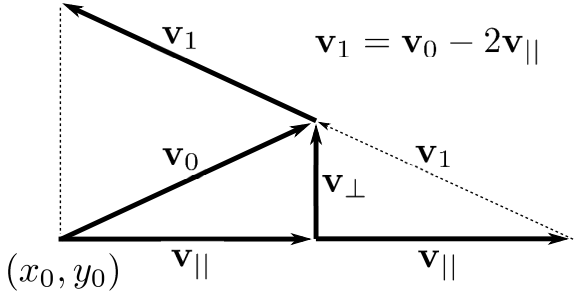}
\caption{\small The ray direction $\bfv_1$ at the opposite branch of the parabola
  after the parabola switching or Snell's reflection.  \label{reflraydir_f4}}
\end{figure}

We intend to avoid tracing the points lying on the parabola. Assume the ray has
already traveled from the point $(x_0,y_0)$ all the way to the axially symmetric
point $(x_1,y_1)$ at the opposite branch. In three-dimensional case its position
and direction have changed to $(\bfr_1,\bfv_1)$. We need to find the increment
$\Delta \bfr$ such that 
\begin{equation}
  \bfr_1 = \bfr_0 + \Delta \bfr 
\end{equation}
Note that $|\Delta \bfr| = \Delta y = 2 (y^* - y_0)$ and $\varepsilon'_0 =
-|\nabla \varepsilon_0|$ because $\varepsilon' < 0$ in the considered vicinity.
Using \eqref{y_extr} yields the length of the increment:
\begin{equation}
  |\Delta \bfr| = \frac{4\varepsilon_0}{|\nabla \varepsilon_0|} 
    \cos \alpha \sin \alpha
\end{equation}
In the chosen $XY$ coordinates $\Delta \bfr$ points in the $Y$ direction.
Decompose $\bfv_0$ into its $X$ and $Y$ components $\bfv_{||}$ and
$\bfv_{\perp}$,
\begin{equation}
  \bfv_0 = \bfv_{||} + \bfv_{\perp}
\end{equation}
Here $|\bfv_0| = 1$, $|\bfv_{||}| = \cos \alpha$, and $|\bfv_{\perp}| = \sin
\alpha$.
We can state that the vector increment $\Delta \bfr$ should be made in the $Y$,
i.e. $\bfv_{\perp}$, direction, so
\begin{equation}
  \Delta \bfr = |\Delta \bfr| \frac{\bfv_{\perp}}{|\bfv_{\perp}|}
\end{equation}
Since $\alpha$ is the angle between $-\nabla \varepsilon_0$ and $\bfv_0$,  
\begin{equation} \label{cosal}
  \cos \alpha = |\bfv_{||}| = \frac{\bfv_0 \cdot (- \nabla \varepsilon_0)}
    {|\nabla \varepsilon_0|},
\end{equation}
whence the vector increment from $\bfr_0$ to $\bfr_1$ at the opposite branch of
the parabola is
\begin{equation} \label{delr0}
  \Delta \bfr = 4 \varepsilon \left( - 
    \frac{\bfv_0 \cdot \nabla \varepsilon_0}
      {|\nabla \varepsilon_0|^2} \right) \bfv_{\perp}
\end{equation}

From \eqref{lineps} the (hypothetical) distance to the critical surface is $L =
1/|\nabla \varepsilon_0|$. With $\cos \alpha$ from \eqref{cosal}, we introduce a
useful expression
\begin{equation} \label{lcosal}
  L \cos \alpha = - \frac{\bfv_0 \cdot \nabla \varepsilon_0}
    {|\nabla \varepsilon_0|^2},
\end{equation}
using which the increment \eqref{delr0} can be written as
\begin{equation} \label{delr}
  \Delta \bfr = 4 \varepsilon L \cos \alpha \bfv_{\perp}
\end{equation}
The new ray direction $\bfv_1$ is calculated from the geometrical considerations
shown in Fig. \ref{reflraydir_f4} as
\begin{equation}
  \bfv_1 = \bfv_0 - 2 \bfv_{||}
\end{equation}

To obtain the vector $\bfv_{||}$, we multiply its absolute value $|\bfv_{||}|$
from \eqref{cosal} by the unity vector along the $X$ axis, which is $- \nabla
\varepsilon_0 / |\nabla \varepsilon_0|$, to get the result:
\begin{equation} \label{v_paral0}
   \bfv_{||} = \frac{\bfv_0 \cdot (- \nabla \varepsilon_0)}
    {|\nabla \varepsilon_0|^2} \nabla \varepsilon_0,
\end{equation}
or, using the \eqref{lcosal} notation,
\begin{equation} \label{v_paral}
   \bfv_{||} = - L \cos \alpha \nabla \varepsilon_0,
\end{equation}
Thus, after the parabola switching, the new ray position and direction are
\begin{subequations} \label{parabl_r1v1}
  \begin{align}
  \bfr_1 &= \bfr_0 + 4 \varepsilon L \cos \alpha (\bfv_0 + 
    L \cos \alpha \nabla \varepsilon_0) \label{parabl_r1} \\
  \bfv_1 &= \bfv_0 + 2 L \cos \alpha \nabla \varepsilon_0 \label{parabl_v1}
  \end{align}
\end{subequations}

\begin{figure}[h]
\epsscale{0.7}
\plotone{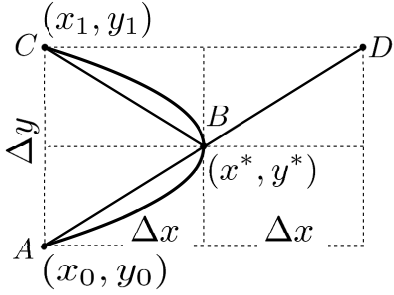}
\caption{\small The length of parabolic segment $ABC$ is approximated by
  the added lengths of two parabolic chords $AB$ and $BC$. This length equals
that of the line segment $AD$. \label{parlenapprox_f5}}
\end{figure}

For the purposes of path integration in line with the trajectory calculation,
the integrand is multiplied by the length of every step $\ud s$. In case of the
parabola switching the integrand is multiplied by the  length of the parabola
segment, $\ud s_{p}$, from $\bfr_0$ to $\bfr_1$. Assuming this length is small
enough it can be replaced by the length of two straight line segments connecting
the points (in $XY$ coordinates) $(x_0,y_0)$, the parabola vertex $(x^*,y^*)$,
and $(x_1,y_1)$, as shown in Fig. \ref{parlenapprox_f5}. Denote as $\Delta x$
the distance from $(x_0,y_0)$ to $(x^*,y^*)$ along the $X$ axis or, which is the
same, along the vector $- \nabla \varepsilon_0$ (see also Fig.
\ref{parrayprm_f3}). Then the parabola length approximation is 
\begin{equation}
  \ud s_{p} = \sqrt{(2 \Delta x)^2 + \Delta y^2}
\end{equation}
We already know that $\Delta y = |\Delta \bfr|$. From \eqref{x_extr} and using
\eqref{lcosal} we can find:
\begin{equation}
   \Delta x = \varepsilon (L \cos \alpha)^2 |\nabla \varepsilon_0| .
\end{equation}

\subsubsection{Linear reflection}

The criterion \eqref{correctness} is based on the linear forward approximation
of $\varepsilon$, and sometimes it works incorrectly if the plasma density
distribution near the critical surface is highly nonlinear (for example,
exponential, as in the chromosphere). Under these conditions the calculated new
point of the ray, $\bfr_1$, appears in the region with $\rho > \rho_{cr}$, which
is impossible in the physical reality. To fix his condition the algorithm stores
in memory the previous ray position and direction $(\bfr_{-1},\bfv_{-1})$. 

When the penetration is detected, the algorithm restores the ray state to one
step back:
\begin{equation}
  (\bfr_0,\bfv_0) \leftarrow (\bfr_{-1},\bfv_{-1}).
\end{equation}
The parabola switching mechanism in this situation can not be relied upon
because the linear Taylor series approximations similar to \eqref{eps0} have
already failed and the parabola parameters would be calculated incorrectly. The
algorithm uses the Newton's method to find the length of step $\ud s$ along the
direction $\bfv_{0}$ such that the next ray point, $\bfr_{1}$, lies on the
critical surface with some small clearance $Tol_{\varepsilon}$. Then the
algorithm performs linear reflection according to the Snell law. Since this
conditions only occur at high frequencies when the rays are close to straight
lines, this procedure does not deteriorate the algorithm's precision.

The Newton's algorithm is based on stepwise approximation:
\begin{equation}
  \ud s_{i+1} = \ud s_i - \frac{\varepsilon_i}
    {\nabla_v\varepsilon_{i}},
\end{equation}
where $\nabla_v \varepsilon_{i} = \bfv_0 \cdot \nabla \varepsilon_{i}$ is the
directional derivative of $\varepsilon_i$ along the vector $\bfv_0$, so the
calculations are performed according to the scheme 
\begin{equation} \label{newton}
  \ud s_{i+1} = \ud s_i - \frac{\varepsilon_i}
    {\bfv_0 \cdot \nabla \varepsilon_{i}},
\end{equation}
The process stops when $|\varepsilon_{i}| \leqslant Tol_{\varepsilon}$. 
The ray direction is then corrected according to the Snell's law. Its geometry
is the same as for the parabola switching shown in Fig. \ref{reflraydir_f4}, so
\begin{equation}
  \bfv_1 = \bfv_0 - 2 \bfv_{||}
\end{equation}
where $\bfv_{||}$ is calculated as in \eqref{v_paral}, and the next algorithmic
step is performed with the last $\ud s$ value immediately from the critical
surface.

\section{Testing and validation}

\begin{figure}[h]
\epsscale{1.0}
\plotone{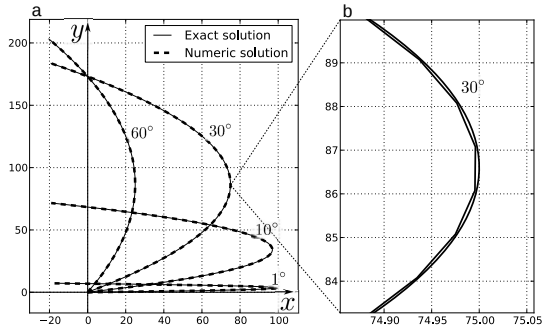}
\caption{\small Comparison of the numerically simulated ray trajectories 
  for several angles of incidence, $1^0, 10^0, 30^0$, and $60^0$, with the exact
ray traces. In panel \textbf{a} the simulated ray curves, shown in thick dashed
lines, overlap the exact parabolic solutions, plotted in solid thin lines. Panel
\textbf{b} shows a closer view of short segments of the $30^0$ traces near the
reflection point at $x = 75$ calculated by the ray tracing algorithm (segmented
line) and the exact solution \eqref{exactsol} (smooth, solid line). Notice the
scales of the axes. \label{numvsexact_f6}}
\end{figure}

The software implementation of the ray tracing algorithm comprises computer
programs in the Fortran 90 and C languages. As an example of the algorithm
validation, the ray trajectories in a hypothetical medium with the linear
dielectric permittivity $\varepsilon$ distribution have been obtained
numerically using the algorithm and compared with the exact solutions. The
results are shown in Fig. \ref{numvsexact_f6}. 

The permittivity $\varepsilon$ linearly falls off along the X axis starting from 1 at $x = 0$ down to 0 at some distance $L$, as given by Eq. \eqref{lineps}. The ray trajectories in this medium have been shown to be parabolas described by Eq. \eqref{pareq}. We chose the distance $L = 100$ of conditional units (they might be solar radii, for example), and set the starting point $\bfr_0$ at the coordinate origin $(0,0,0)$. Here the ray equation \eqref{pareq} takes the form
\begin{equation} \label{exactsol}
  x = L \cos^2 \alpha - L \left( \cos \alpha - 
    \frac{1}{2 L \sin \alpha} y \right)^2, 
\end{equation}
where $\alpha$ is the angle of incidence on the critical surface at $x = L$. The
results of comparison are presented in panel \textbf{a} of Fig.
\ref{numvsexact_f6}. Four angles $\alpha$ were selected: $1^0, 10^0, 30^0$, and
$60^0$. The exact solutions calculated by \eqref{exactsol} are plotted in thin
solid lines; the ray traces created by the algorithm are plotted over them as
thicker dashed lines. The dashed curves cover the solid ones, so one can
conclude the numerical solutions are essentially precise. In panel \textbf{b} of
Fig. \ref{numvsexact_f6} the differences between the numerical and exact
solutions are more discernible because only a short piece of the trajectory in
the vicinity of the parabola vertex is shown.

\section{Example applications}

The algorithm has been applied for solar studies and shown good results. A
series of numerical experiments was conducted to simulate the brightness
temperature distribution, $T_B$, across the solar disk as seen from the earth,
in the wide frequency range from 10 MHz to 3 GHz. The calculation of $T_B$
requires the ray tracing. We used different electron density distributions,
$N_e$ ($\mathrm{cm}^{-3}$), for the corona and the chromosphere. The coronal
$N_e$ was approximated by the \citet{sai70} model:
\begin{align} \label{Saito}
N_e(r,\theta) &= 3.09 \times 10^8 r^{-16} (1 - 0.5\cos \theta) + \nonumber \\ &
  1.56 \times 10^8 r^{-6} (1 - 0.95\cos \theta) + \nonumber \\ & \quad 0.0251
\times 10^8 r^{-2.5} (1 - \sqrt{\cos \theta}),
\end{align}
where $r$ is distance in solar radii, $R_{\odot}$, from the sun center and
$\theta$ is the heliographic colatitude. In the chromosphere, up to 9,000 km
from the photosphere, we used the \cite{cme35} model:
\begin{equation} \label{Menzel}
N_e(r) = 5.7 \times 10^{11} e^{-7.7 \times 10^{-4}(R_{\odot}(r-1) - 500)},
\end{equation}
and in the layer between the chromosphere and corona, for the altitudes from
9,000 km to 11,000 km,  a smooth polynomial patch function was used.

\begin{figure}[h]
\epsscale{1.0}
\plotone{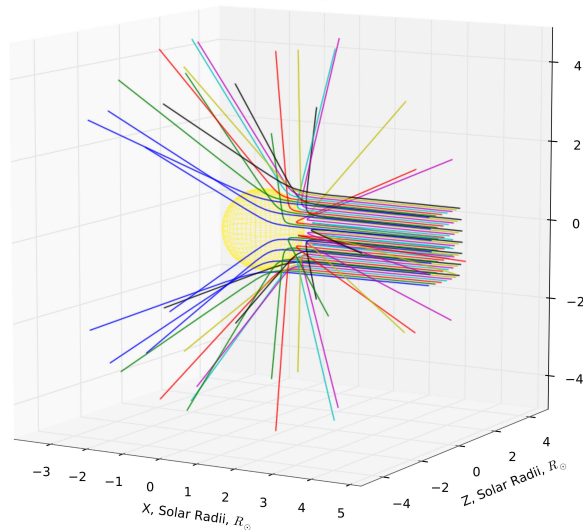}
\caption{\small The result of numerical simulation of a beam of electromagnetic ray traces at 80 MHz in the solar corona. The square cross-section beam consists of $7 \times 7$ evenly spaced rays directed along the $X$ axis in the $HEE$
coordinate system \citep{frh02}, i.e. from the earth. The sun is shown as a
yellow sphere.    \label{raybunch_f7}}
\end{figure}

In Fig. \ref{raybunch_f7} is shown a bunch of the computed ray trajectories at
80 MHz, refracting in the coronal and chromospheric plasmas. The $7 \times 7 =
49$ evenly spaced rays comprise a beam converging to a point at the earth
distance (here 215 $R_\odot$), where the rays can be displayed as a $7 \times
7$-pixel raster image. Note that in our experiments with the ray-tracing
algorithm the typical raster sizes were from $100 \times 100$ to $1000 \times
1000$ and more; the beam in Fig. \ref{raybunch_f7} is shown for illustrative purposes only. As seen, the rays arriving at the earth to form an image at 80 MHz can carry information from very different heliospheric regions due to differences in refraction across the image plane. 

\begin{figure}[h!]
\epsscale{1.0}
\plotone{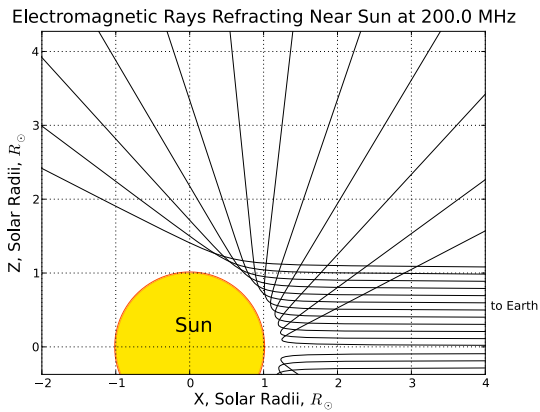}
\caption{\small Refraction in the corona at 200 MHz in the $XZ$ plane of the $HEE$
coordinate system \citep{frh02}. The $X$ points at the earth, the $Z$ is normal
to the ecliptic plane. The set of propagating rays is shown in thin solid lines.
The zone of avoidance close to the solar surface, outlined by the envelope of
rays, is clearly  visible at $~1.2 R_{\odot}$.  \label{refrac200mhz_f8}}
\end{figure}

Fig. \ref{refrac200mhz_f8} shows the results of ray tracing at 200 MHz in the
$XZ$ plane of the the $HEE$ \citep{frh02} coordinate system (the $X$ points from
the sun center at the earth). The rays make up an envelope outlining the
critical surface, which is at approximately $1.2 R_\odot$ from the sun center. 

\begin{figure}[h]
\epsscale{1.0}
\plotone{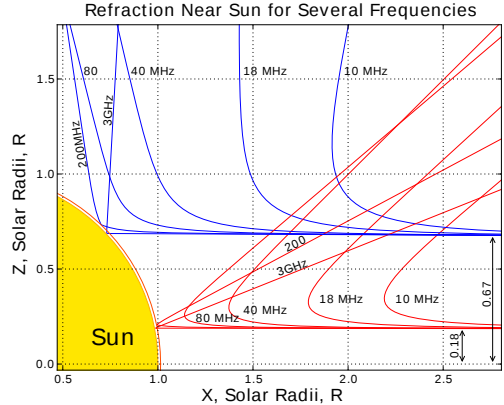}
\caption{\small Refraction near the sun for several frequencies (``spectral
decomposition''). Two thin beams, each composed of the frequencies 10, 18, 40,
80, 200 MHz, and 3 GHz, are directed along the $X$ axis in the $XZ$ plane in the
$HEE$ coordinates \citep{frh02}. The $X$ points at the earth, the $Z$ is normal
to the ecliptic plane. The beams' axes are at $z = 0.18 R_{\odot}$ and z = $0.67
R_{\odot}$. One can see the observations at different frequencies are dominated
by different layers of the corona.  \label{refracfreq_f9}}
\end{figure}

The higher the frequency, the deeper the electromagnetic rays penetrate into the
solar vicinity. Fig. \ref{refracfreq_f9} helps estimate the solar distances
where the rays at different frequencies start to significantly change the
direction. The two thin beams of rays are directed parallel to the $HEE$ $X$
axis at the $Z$ distances of $0.18 R_\odot$ and $0.67 R_\odot$, respectively.
Each beam includes rays at the frequencies 10, 18, 40, 80, 200 MHz, and 3 GHz.
The Fig. \ref{refracfreq_f9} can be also viewed as the ``spectral
decomposition''. One can notice that at 3 GHz the rays are virtually straight
lines.

\section{Conclusions}

The proposed algorithm is intended for efficient computation of electromagnetic
ray traces in the heliospheric plasmas in parallel with integration over the ray
paths. To be fast, it is based on a difference scheme of the second order of
precision, but it has a property of conserving the direction cosine vector
length of the computed trajectory. With this benefit, the algorithm maintains
sufficient precision. The difference scheme is only a part of the algorithm:
besides, it incorporates several mechanisms for adaptive estimation of the
optimal step and prevention of occasional ray penetrations into the regions
where the plasma frequencies are higher than the own ray frequency. 

The algorithm has been tested on problems of the solar disk image simulation,
and it demonstrated good speed, precision, and robustness. We presented here
some results pertaining to the ray traces calculation only; the main results
will be published in a separate paper. However, we can conclude that, as seen
from Fig. \ref{refracfreq_f9}, observations at different frequencies are
dominated by different layers of the corona and chromosphere The broad band
observations could help reveal the large scale structures in these domains,
which can be likened to peeling the layers of an onion bulb. The new
high-resolution radiotelescope MWA with its frequency range from 80 to 300 MHz
therefore will be an effective instrument in the solar studies. The algorithm
and software described in this paper were intended for use as the analysis tool
for the MWA solar images. 

However we should specially point out that the algorithm applications are in no
way confined to the MWA project only. The algorithm can be used for virtually
any radioastronomical study because of its very wide frequency range. 
 
Among possible future improvements to the algorithm we could mention the plans
on employment of the multi-core, hyperthreading, and multi-channel-memory powers
of the modern computers. The problem of simultaneous computation of many ray
trajectories can be effectively decomposed into several threads, each running on
a separate CPU, because the rays in a beam are calculated independently.  

\acknowledgments
\textbf{ACKNOWLEDGEMENTS}

This work was supported by a NSF grant to the MIT Haystack Observatory. MWA is a
large international collaboration, the partner institutions for which are listed
below. The US part of the MWA collaboration is funded by the NSF Astronomy and
Atmospheric Sciences Divisions.

\clearpage

\end{document}